\newcommand{\bee} {\begin{equation}}
\newcommand{\ene} {\end{equation}}
\newcommand{\beqa} {\begin{eqnarray}}
\newcommand{\enqa} {\end{eqnarray}}
\newcommand{\beqsa} {\begin{eqnarray*}}
\newcommand{\enqsa} {\end{eqnarray*}}
\newcommand{\bea} {\begin{array}}
\newcommand{\ena} {\end{array}}

\documentstyle[12pt]{article}
\begin{document}
\baselineskip=18pt

{{\hfill FERMILAB-Pub-93/380-A}}\\
\bigskip
{\hfill December 1993}

\begin{center}
{\Large\bf Quantum statistical effects on parton distribution scaling
behaviour}\\
\end{center}

\bigskip\bigskip

\begin{center}
{\bf G. Mangano$^{a,b}$, G. Miele$^{b,c,*}$} and
{ \bf G. Migliore$^a$}

\end{center}

\bigskip

\small
{\it ${}^a$ ~~Dipartimento di Scienze Fisiche, Universit\`a
di Napoli - Federico II -, Mostra D'Oltremare Pad. 19, 80125, Napoli, Italy.}

{ \it ${}^b$~~INFN, Sezione di Napoli, Mostra D'Oltremare Pad. 20,
80125, Napoli, Italy.}

{\it  ${}^c$~~NASA/Fermilab Astrophysics Center, Fermilab National
Accelerator Laboratory, Box 500, Batavia, IL60510-0500, USA.}
\normalsize

\bigskip\bigskip\bigskip

\begin{abstract}

Starting from the old idea that Fermi statistics for quarks play
a fundamental role to explain  some features of hadron structure,
we study the modification of the scaling behaviour of parton distributions due
to quantum statistical effects. In particular, by following an interesting
formal analogy which holds between the Altarelli-Parisi evolution equations,
in leading-log approximation, and a set of Boltzmann equations,
we generalize the evolution equations to take into
account Pauli exclusion principle and gluon induced emission.

\end{abstract}

\vspace{2cm}

\centerline{{\normalsize \it Submitted to Phys. Lett. B}}

\vspace{2cm}

\noindent

\footnotesize

${ }^{*}$On leave of absence from the Dipartimento di Scienze Fisiche,
Universit\`a di Napoli, Italy.
\normalsize
\newpage

Deep inelastic experiments seem to be an inexhaustible source of information
on the hadronic structure and continue to considerably improve our
understanding of strong interaction dynamics. A measurement of proton and
neutron $F_{2}(x)$ structure function performed by the NMC Collaboration at
CERN \cite{nmc} suggests a rather large $SU(2)$ flavour breaking in the sea
quark \cite{prep}. In particular they have obtained a determination,
at very small $x$, for the difference
\bee
{\cal I}_{G}(x) = \int_{x}^{1} \frac {dy}{y} [ F_{2}^{\mu p}(y)-
F_{2}^{\mu n}(y)]
= \frac{1}{3} \int_{x}^{1} dy~ [ u(y)+{\bar u}(y)-d(y)-{\bar d}(y)]~~~,
\label{eq:gsr}
\ene
finding ${\cal I}_{G}(0.004) = 0.227 \pm 0.007$ .
Thus, by extrapolating down to $x = 0$, they have estimated
\bee
{\cal I}_{G}(0) = 0.240 \pm 0.016~~~.
\label{eq:ig0}
\ene
This result represents a relevant violation of the Gottfried sum rule
\cite{got}, which would predict ${\cal I}_{G}(0) = 1/3$.
Moreover, from (\ref{eq:ig0}) we get
\bee
\bar{d}-\bar{u}=\int_{0}^{1} dx~ [ \bar{d}(x)- \bar{u}(x) ] \sim 0.14~~~.
\label{eq:du}
\ene
However, the inequality $\bar{d}>\bar{u}$ was already argued many years ago
by Field and Feynman \cite{ff} on pure statistical basis. They suggested
that in the proton the production from gluon decays of $u \bar{u}$-pairs
with respect to $d \bar{d}$-pairs would be suppressed by Pauli principle
because of the presence of two valence $u$ quarks but of only one valence $d$
quark. Assuming this point of view, the experimental result (\ref{eq:ig0})
naturally leads to the conclusion that quantum statistical effects play
a sensible role in parton dynamics and that, in particular, parton
distribution functions are affected by them. In this picture one may
also easily account for the known dominance at high $x$ of $u$-quarks
over $d$-quarks, whose characteristic signature is the fast decreasing of the
ratio $F_{2}^{n}(x)/F_{2}^{p}(x)$ in this regime.

In recent papers \cite{buc} this idea has been developed, succeeding in making
reasonable assumptions for various polarized parton distributions in terms of
unpolarized ones, explaining the observed violation of Ellis-Jaffe sum rule
\cite{ej}, and giving a possible solution to the spin crisis problem
\cite{emc}. Finally,
using Fermi-Dirac or Bose-Einstein inspired form for parton distribution
functions, a rather good agreement has been obtained with the experimental
data on structure functions \cite{bour}.

The aim of this letter is to study the role of quantum statistical effects,
namely Pauli exclusion principle and induced gluon emission, in the
$Q^2$-evolution
of parton distributions. We will show that there is a quite close
analogy between the well-known Altarelli-Parisi (A-P) evolution equations
\cite{ap}, in leading-log approximation, and a set of Boltzmann equations
written for a dilute system of particles. This analogy will guide us in
finding the generalized scaling when also quantum statistics have been taken
into account.

As well-known, the logarithmic dependence on $Q^2$ of the parton distribution
momenta, predicted in the framework of perturbative $QCD$, has a simple and
beautiful interpretation in terms of evolution equations for parton
distribution functions. At leading-log level, the A-P equations can be written
in the following way
\bee
{d \over {dt}}p_{A}(x,t)= {\alpha_{s}(t) \over 2 \pi} \int_{x}^{1} {dy \over y}
\sum_{B} p_{B}(y,t) P_{AB}\left( {x \over y} \right)~~~,
\label{eq:ap}
\ene
where $t = \ln(Q^{2}/\mu^{2})$, $\mu$ is some renormalization
scale and $p_{A}(x,t) $  denote the parton distribution functions
($A,B=$quarks, antiquarks and gluons). By defining
\bee
\tau \equiv \frac{1}{2 \pi b} \ln \left[ \frac{\alpha(\mu^{2})}{\alpha(t)}
\right]~~~,
\label{eq:tau}
\ene
with $b \equiv (33-2n_{f})/(12 \pi)$ ($n_{f}$ is the number of flavours),
Eq. (\ref{eq:ap}) becomes
\bee
{d \over {d\tau}}p_{A}(x,\tau)= \int_{x}^{1} {dy \over y}
\sum_{B} p_{B}(y,\tau) P_{AB}\left( {x \over y} \right)~~~.
\label{eq:apb}
\ene
Note that the dependence on $\tau$ of r.h.s. of (\ref{eq:apb}) comes only
through $p_{B}(y,\tau)$.\\ In Eqs. (\ref{eq:ap}) and (\ref{eq:apb}),
$P_{AB}(x/y)$ stand for the
splitting functions evaluated by using standard {\em equivalent parton
method}. They correspond to the probability for the elementary three-body
processes to occur in which a parton with momentum fraction $x$ is produced
by a parton with higher fraction $y=x/z$.

The simple microscopical interpretation of Eq. (\ref{eq:apb})
is, as well-known, that the $\tau$ dependence of $p_{A}$ distributions
is induced
by these processes considering them as occurring in the vacuum.\\
However, it is physically reasonable to imagine that
this picture has to be modified
for sufficiently low $x$; in this regime
the nucleons are filled with a large number of quark-antiquark
pairs and gluons (the sea) and thus,
to take into account in
the correct way the presence of this large number of partons,
the decays $A \rightarrow B+C$ should
be considered in presence of a surrounding plasma of both
Fermi and Bose particles.
Corrections induced by quantum statistical effects to the scaling
behaviour dictated by (\ref{eq:apb}) are therefore generally
present, and in particular we expect that:
\begin{itemize}
\item[a)] Pauli blocking will suppress the production of quarks and
antiquarks with fraction $x$ corresponding to $filled$ levels;

\item[b)] the gluon emission probability through bremsstr\"{a}hlung
processes, considered in the standard picture leading
to A-P equations, will be enhanced by the contribution
of induced-emission in presence of a rather relevant
number of gluons in the sea.
\end{itemize}
These effects would favour the production of gluon-quark pairs with
larger values of $x$ for the quarks and a smaller one for the gluon.
Moreover the gluon conversion processes in $q-\bar{q}$ pairs are expected to
be reduced.

In statistical mechanics all these effects are simply included by multiplying
the amplitudes modulus squared of the relevant processes by the factors $1-f$
or $1+f$ for each Fermi or Bose particle in the final state, with $f$
denoting the particle distribution functions without any level-density
factor. In equilibrium conditions these $f$ reach the standard stationary
Fermi-Dirac or Bose-Einstein form, while in general they depend on time.
Thus, it is reasonable to expect that similar factors should be introduced in
the A-P equations.

To go further on this point we will show, as already mentioned, that
A-P evolution equations can be formally viewed as Boltzmann transport
equations for parton distributions when they satisfy classical statistics
(dilute system).

As well-known, the Boltzmann set of equations describes the evolution to
equilibrium states of systems composed by many particles of several species
($i$ specie-index $i = 1,..,n$)
mutually interacting \cite{degr}. Assuming, for the following applications,
monodimensional dynamics for particles  we can define the numerical
distribution functions as
\bee
n_{i}(\epsilon, t) \equiv g_{i}(\epsilon)f_{i}(\epsilon,t)~~~,
\label{eq:ni}
\ene
with $\epsilon$ denoting the energy, $f_{i}(\epsilon,t)$ the statistical
functions (they recover the usual Bose/Einstein or Fermi/Dirac at the thermal
equilibrium), and $g_{i}(\epsilon)$ the level-densities (weights)
corresponding to $\epsilon$. These last quantities should be fixed from the
beginning,
by studying the hamiltonian of the total system. From (\ref{eq:ni}) follows
the expression for the total number of i-particles
\bee
N_{i}(t) = \int d\epsilon~g_{i}(\epsilon)f_{i}(\epsilon,t)~~~.
\label{eq:nit}
\ene
By using Eq. (\ref{eq:ni}), the Boltzmann equations can be cast in
the following form
\bee
{\cal L}~ n_{i} = C_{i}[{\bf f},{\bf g}] = C^{+}_{i}[{\bf f },{\bf g}]-
C^{-}_{i}[{\bf f},{\bf g}]~~~~~~~~~~~~~~~~~i=1,....,n~~~~,
\label{eq:liouv}
\ene
where ${\bf f} \equiv (f_{1}, ..., f_{n})$, ${\bf g} \equiv (g_{1}, ...,
g_{n})$, ${\cal L}$ is the Liouville operator, and $C_{i}[{\bf f},{\bf g}
]$ is the so
called collisional integral for the i-th  particle specie. The latter
is given by a thermal average of all possible processes which
change the density of the i-th specie.
Notice that in Eq. (\ref{eq:liouv}) we have defined $C^{+}_{i}[{\bf f},{\bf g}
]$ and $C^{-}_{i}[{\bf f},{\bf g}]$  as the contributions corresponding to the
interaction processes which create or destroy the i-th particle specie
respectively.
For  simple three body processes $A \rightarrow B+C$, $B \rightarrow A+C$,
if we are interested in describing the modification of $B$ population,
the corresponding terms in $C_{B}[{\bf f},{\bf g}]$ are the following
\beqa
C^{+}_{B}[{\bf f},{\bf g}] - C^{-}_{B }[{\bf f},{\bf g}]  =
2 \pi~\int \int d\epsilon_{A}~ d\epsilon_{C}~\Bigr\{ |{\cal M}(A \rightarrow
B+C)|^2
{}~\delta(\epsilon_{A}-\epsilon_{B}-\epsilon_{C})\nonumber \\
\times~
n_{A}(\epsilon_{A},t) ~g_{B}(\epsilon_{B})~\left[1 \pm f_{B}(\epsilon_{B},t)
\right]~ g_{C}(\epsilon_{C})~\left[1 \pm f_{C}(\epsilon_{C},t)\right] \Bigr\}
\nonumber \\
 - 2 \pi~ \int \int d\epsilon_{A}~d\epsilon_{C}
{}~\Bigr\{ |{\cal M}(B \rightarrow A+C)|^2
{}~\delta(\epsilon_{B}-\epsilon_{A}-\epsilon_{C})\nonumber \\
\times~
n_{B}(\epsilon_{B},t) ~g_{A}(\epsilon_{A})~\left[1 \pm f_{A}(\epsilon_{A},t)
\right]~ g_{C}(\epsilon_{C})~\left[1 \pm f_{C}(\epsilon_{C},t)\right] \Bigr\}
\nonumber\\
\label{eq:coll}
\enqa
where  ${\bf f}
\equiv(f_{A}(\epsilon_{A},t),f_{B}(\epsilon_{B},t),f_{C}(\epsilon_{C},t))$,
${\bf g} \equiv (g_{A}(\epsilon_{A}),g_{B}(\epsilon_{B}),g_{C}
(\epsilon_{C}))$,
$|{\cal M}|^2$ are the squared
moduli of transition amplitudes and the sign in the final state factors is
positive/negative depending on the bosonic/fermionic nature of particles.
In the limit of very small $f_{i}$ one has $(1\pm f_{i}) \sim 1$,
and assuming free particle states ($g_{i}=const$)
the collisional term for very dilute systems is recovered.

Assuming $n_{f}$ different flavours for quarks
($j=1,...,n_{f}$) with elicity states ($\lambda=+,-$), we
can rewrite the set of equations (\ref{eq:ap})
for polarized quarks $(q_{j \lambda})$,
antiquarks $(\bar{q}_{j \lambda})$, and not-polarized  gluons $(G)$
distribution functions in terms of two-dimensional integrals
\begin{eqnarray}
\frac{d}{d\tau} q_{j\lambda}(x,\tau)  & = &  \int_{0}^{1}\int_{0}^{1}
dy dz ~\delta(x-yz)~[P_{qq}(z) q_{j\lambda}(y,\tau) + {1 \over 2}
P_{qG}(z) G(y,\tau)]~~~,
\label{eq:bidima}\\
\frac{d}{d\tau}
\bar{q}_{j\lambda}(x,\tau) & = &  \int_{0}^{1}\int_{0}^{1}
dy dz ~\delta(x-yz)~[P_{qq}(z)
\bar{q}_{j\lambda}(y,\tau) +  {1 \over 2} P_{qG}(z) G(y,\tau)]~~~,
\label{eq:bidimc}\\
\frac{d}{d\tau} G(x,\tau)  &=&  \int_{0}^{1}\int_{0}^{1}
dy dz~ \delta(x-yz)~\Bigr\{P_{GG}(z) G(y,\tau)\nonumber\\
& + & \sum_{j=1}^{n_{f}} \sum_{\lambda=
+,-}P_{Gq}(z)\Bigr[ q_{j\lambda}(y,\tau)+\bar{q}_{j\lambda}(y,\tau)
\Bigr] \Bigr\}~~~.
\label{eq:bidimb}
\end{eqnarray}
We have assumed for polarized gluon distributions that
$G_{+}(x,\tau) \sim G_{-}(x,\tau) \sim G(x, \tau)/2$:
we will comment on this point later.
Note that in the previous expressions the integrating-variables $y$ and $z$
vary from $0$ to $1$, i.e. to the maximum available energy properly normalized.
Starting from the first of above equations,
Eq. (\ref{eq:bidima}), we notice that
if we formally regard $\tau$ as a {\it time} parameter,
r.h.s. of (\ref{eq:bidima}) represents the $C^{+}_{j\lambda}$
collisional term of a Boltzmann equation,
written for particles obeying to a
monodimensional dynamics. This equation is given in terms
of numerical distribution functions $q_{j\lambda}(x,\tau)$, $G(x,\tau)$ and of
the probabilities for the elementary processes $P_{qq}(z)$ and $P_{qG}(z)$
(the final particles are assumed to be free which means no presence of
extra-$g$ terms corresponding to them).
It is worth-while pointing out that in the infinite-momentum frame,
where the parton
picture is well-defined, all transverse dynamics can be safely
neglected (it has been already integrated out, indeed) and thus the
description is monodimensional.
The $\delta$-function in
(\ref{eq:bidima}) is just the longitudinal momentum conservation in the
three body interaction process.\\
However, to complete the analogy between
(\ref{eq:bidima}) and the corresponding Boltzmann equation we still
have the difficulty that no $C^{-}_{j\lambda}$ terms are included
in (\ref{eq:bidima}): these
correspond to decays of the $x$-momentum quark in parton pairs and would only
depend on $ q_{j\lambda}(x, \tau)$ if all
statistical factors for final states are
neglected. The splitting functions
 involved in this case are two:
$P_{qq}(z)$ and $P_{Gq}(z)$, being respectively, the emission probability of a
quark or a gluon with fraction $y=zx$. Notice that in $C^{-}_{j\lambda}$
terms the
momentum conservation leads to a definition of $z$ which is simply the inverse
of the one in $C^{+}_{j\lambda}$. Thus, the $C^{-}_{j\lambda}$
collisional integral would be
\bee
C^{-}_{j\lambda}\left[{\bf q}(x,\tau),G(x,\tau)\right]= q_{j\lambda}
(x, \tau) \int_{0}^1
\int_0^1 dy dz~ \delta \left( x- {y \over z} \right) ~[P_{qq}(z) +
P_{Gq}(z)]~~~,
\label{eq:inv}
\ene
with ${\bf q}=(q_{1+},..,q_{n_{f}+},q_{1-},..,q_{n_{f}-})$, and
by integrating over $y$ we finally get
\bee
C^{-}_{j\lambda}
\left[{\bf q}(x,\tau),G(x,\tau)\right] = q_{j\lambda}(x, \tau) \int_0^1 dz
{}~z ~ [ P_{qq}(z) + P_{Gq}(z) ]= 0~~~,
\label{eq:zero}
\ene
where in (\ref{eq:zero}) we have used the integral constraint following from
momentum conservation in quark splitting. Therefore from (\ref{eq:zero})
since the $C^{-}_{j\lambda}$ contribution vanishes, we
deduce that A-P evolution equation for $q_{j\lambda}$ distribution can be
consistently viewed as a Boltzmann equation in which all statistical factors
corresponding to final particle are neglected (dilute system).
Similar considerations
can be repeated for the antiquarks evolution equations
(\ref{eq:bidimc}).

A completely analogous result also holds for gluon distribution, whose
evolution is dictated by (\ref{eq:bidimb}). In this case,
by writing the corresponding Boltzmann {\it inspired} equation, one gets
\beqa
{ d \over { d \tau} } G(x, \tau)  =  \int_0^1 \int_0^1 dy ~dz ~\delta( x-yz)
\Bigr\{P_{GG}(z)G(y, \tau)
 +  \sum_{j=1}^{n_{f}} \sum_{\lambda=+,-} P_{Gq}(z)
\nonumber\\ \times
 \Bigr[ q_{j\lambda}(y, \tau) + \bar{q}_{j\lambda}(y, \tau)
\Bigr] \Bigr\}
 - G(x, \tau) \int_0^1 \int_0^1  dy ~dz ~ \delta \left( x - { y \over z}
\right) [ P_{GG}(z) + 2n_{f} P_{qG}(z)] \nonumber \\
 =  \int_x^1 { dy \over y} \left\{ P_{GG} \left( {x \over y} \right) G(y,
\tau) + \sum_{j=1}^{n_{f}} \sum_{\lambda=+,-}
  P_{Gq} \left( { x \over y } \right)\Bigr[ q_{j\lambda}(y, \tau) +
\bar{q}_{j\lambda}(y, \tau) \Bigr] \right\},
\label{eq:gluons}
\enqa
which is just the A-P equation for gluon distributions. The $C^{-}_G$
term still vanishes due to the momentum conservation constraint
\bee
\int_0^1 dz~z [ P_{GG}(z) + 2 n_{f} P_{qG}(z)] = 0~~~.
\label{eq:zerogluons}
\ene
Let us briefly summarize our results till this point: we have
shown that A-P evolution
equations can be formally regarded as a set of Boltzmann equations for parton
distribution functions, in which the Liouville operator takes the simple form
of derivative with respect to $\tau$ scaling variable. It is worth to notice
that the absence of any
external force in the regime of high transferred $Q^2$ is of course
compatible
with this expression for ${\cal L}$. The analogy holds under the hypothesis
that quarks and gluons form a very dilute system in the nucleons, so the
statistical
factors for final particles in the interaction processes can be neglected.
Starting from this equivalence and urged from the idea that, instead, quantum
statistics would play a role in parton dynamics, it is now easy to generalize
A-P equations to a set of generalized scaling equations where Pauli exclusion
principle and gluon stimulated emission processes can be taken into account
in a simple way.

To this aim, as in equation (\ref{eq:coll}), one should introduce in the
collisional integrals the $(1 \pm f_{i})$ factors, and thus, as long as the
statistical effects are taken into account, the factorization of
$q_{j\lambda}$, $\bar{q}_{j\lambda}$ and
$G$ as reported in Eq. (\ref{eq:ni}) becomes necessary.\\
In the same spirit of (\ref{eq:ni}), we will write the quark, antiquark and
gluon distributions as
\begin{eqnarray}
q_{j\lambda}(x, \tau) & = & g_{j\lambda}(x)~f_{j}^{\lambda}(x, \tau)~~~,
\label{eq:prod1}\\
\bar{q}_{j\lambda}(x, \tau) & = &
\bar{g}_{j\lambda}(x)~\bar{f}_{j}^{\lambda}(x, \tau)~~~,\label{eq:prod2}\\
G(x, \tau) & = & g_{G}(x)~f_{G}(x, \tau)~~~,
\label{eq:prod3}
\end{eqnarray}
where $g_{j\lambda}(x)$, $\bar{g}_{j\lambda}(x)$ and $g_{G}(x)$
are weight functions, whereas
$f_{j}^{\lambda}(x,\tau)$, $\bar{f}_{j}^{\lambda}(x,\tau)$
and $f_{G}(x,\tau)$ are purely statistical distributions.
The explicit form for $g$-functions, which contains
the infrared divergency at
$x=0$, should be fitted from experimental data, as in \cite{bour}, or deduced
from theoretical expected behaviour, like, for example, Regge theory.
We stress that the factorized form (\ref{eq:prod1})-(\ref{eq:prod3}),
in particular the
hypothesis that the singular functions $g_{j\lambda}$, $\bar{g}_{j
\lambda}$ and
$g_{G}$ do not depend on $\tau$ is compatible with
predictions of both Regge theory and $QCD$ for the behaviour of parton
distributions at the end-point $x=0$; as well-known in this regime one has
\bee
p_{A}(x, Q^2) \sim \xi_{A}(Q^2) x^{- \alpha_{A}}~~~,
\label{eq:regge}
\ene
with $\alpha_{A}$ which does not depend on $Q^2$, at least for large $Q^2$
\cite{ynd}.\\
Remarkably, a parameterization similar to (\ref{eq:prod1})-(\ref{eq:prod3})
has been already successfully proposed on
phenomenological basis
in \cite{bour}, to fit all the available measurements on parton
distributions at fixed $Q^2$, assuming for them
a thermal-equilibrium form
\begin{eqnarray}
q_{j\lambda}(x) &=&A~x^{-\alpha}~
\left[\exp\left( { { x - {x}_{j\lambda}} \over
\tilde{x} } \right) + 1 \right]^{-1}~~~,\label{eq:prop1}\\
\bar{q}_{j\lambda}(x) &=&A~x^{-\alpha}~
\left[\exp\left( { { x - \bar{x}_{j\lambda}} \over
\tilde{x} } \right) + 1 \right]^{-1}~~~,\label{eq:prop2}\\
G(x) &=& { 16 \over 3 }~A~x^{-\alpha}~\left[\exp\left( { { x - {x}_{G}} \over
\tilde{x} } \right) - 1 \right]^{-1}~~~,
\label{eq:prop3}
\end{eqnarray}
where ${x}_{j\lambda}$, $\bar{x}_{j\lambda}$, and ${x}_{G}$
represent the {\it thermodynamical potentials}, and
$\tilde{x}$ plays the role of the {\it temperature}.
It is worth-while to point out that, in the framework of a
formal connection between
the A-P evolution equations and Boltzmann equations, these results \cite{bour}
are a straight consequence of the above analogy, which predicts
thermal-equilibrium-like solutions for parton distribution functions
for sufficiently high values of $Q^2$.

Within the factorized expression (\ref{eq:prod1})-(\ref{eq:prod3})
the final state factors
are written in the form $ 1 - f_{j}^{\lambda}$, $ 1 - \bar{f}_{j}^{\lambda}$
and $ 1+f_{G}$ for quarks, antiquarks and gluons respectively.

We are now able to introduce a set of generalized scaling equations for
quarks and gluons.
% when the final particles are produced as free (the
%corresponding density-states equal to $1$ in the Bijorken-variable space).
Here we will consider for simplicity
the case in which the gluons are supposed to
not have a significant net polarization in the nucleons with respect to the
one carried by quarks. We will assume, therefore $ G_{+}(x, \tau) \sim
G_{-}(x, \tau) \sim G(x, \tau)/2$.
It should be pointed out that this approximation is consistent with the
results obtained in \cite{buc} and \cite{bour}, where it is argued that still
Pauli principle plays an essential role to generate the polarization of the
quark sea, in the same approach therefore adopted here. \\
It is, instead, less
satisfactory in the framework of the different interpretation of the violation
of Ellis-Jaffe sum rule based on the
axial-vector current anomaly \cite{anom}.
This latter case, in fact, would require a very large
gluon polarization, i.e. $\Delta G = G_{+} - G_{-} \sim 3 \div 4$.
Notice however that, as shown in \cite{bour}, gluons are expected to be more
numerous than quarks, due to their Bose nature, so in any case one has
$\Delta G /G << \Delta q/q$, which supports our approximation.

By helicity conservation at the quark-gluon vertex, it is easily seen that
evolution equations for polarized quark distribution functions get the
following form
\beqa
 { d \over { d \tau} } q_{j\lambda}(x, \tau) & = & \int_x^1 { dz \over z}
\left\{ P_{qq}(z)~ q_{j\lambda} \left( { x \over z }, \tau \right)
\left[ 1 -  f_{j}^{\lambda} (x, \tau)\right]
\left[ 1 +  { 1 \over 2 } f_{G} \left( x \left(
 { 1 \over z} -1 \right), \tau \right) \right] \right. \nonumber \\
& +& \left.  { 1 \over 2 } P_{qG}(z)~ G \left( { x \over z }, \tau \right)
\left[ 1 -  f_{j}^{\lambda} (x, \tau)\right]
\left[ 1 -  \bar{f}_{j}^{-\lambda} \left( x \left(
{ 1 \over z} -1 \right), \tau \right) \right] \right\}  \nonumber \\
&- &  q_{j\lambda}(x, \tau) \int_0^1 z~ dz \left\{ P_{qq}(z)~
\left[ 1 -  f_{j}^{\lambda} (xz, \tau)\right]
\left[ 1 +  { 1 \over 2 } f_{G} \left( x \left(
1 - z \right), \tau \right) \right]  \right. \nonumber \\
& +& \left. P_{Gq}(z)~
\left[ 1 +  { 1 \over 2 } f_{G} (xz, \tau) \right]
\left[ 1 -  f_{j}^{\lambda} \left( x \left(
1 -  z \right), \tau \right) \right] \right\}~~~.
\label{eq:quark}
\enqa
The equations for antiquarks are easily obtained by the previous one by
substituting $q_{j\lambda} \leftrightarrow \bar{q}_{j \lambda}$ and
$f_{j}^{\lambda} \leftrightarrow \bar{f}_{j}^{\lambda}$.
Similarly for the gluon distribution $G(x,\tau)$ one has
\beqa
{ d \over { d \tau} } G (x, \tau) &=& \int_x^1 { dz \over z}
\left\{ P_{GG}(z)~ G \left( { x \over z }, \tau \right)
\left[ 1 +  { 1 \over 2 } f_{G} (x, \tau)\right]
\left[ 1 +  { 1 \over 2 } f_{G} \left( x \left(
{ 1 \over z} - 1 \right), \tau \right) \right] \right. \nonumber \\
& +& \sum_{j=1}^{n_{f}} \sum_{\lambda=+,-} P_{Gq}(z)~
\left[ 1 +  { 1  \over 2 } f_{G} (x, \tau) \right]
\left\{ q_{j\lambda} \left( { x \over z }, \tau \right)
\left[ 1 -  f_{j}^{\lambda} \left( x \left(
{ 1 \over z} -1 \right), \tau \right) \right] \right.
\nonumber\\
& + & \left. \left. \bar{q}_{j\lambda} \left( { x \over z }, \tau \right)
\left[ 1 -  \bar{f}_{j}^{\lambda} \left( x \left(
{ 1 \over z} -1 \right), \tau \right) \right]
\right\} \right\}
 \nonumber \\
& - &  G(x, \tau) \int_0^1 z~ dz \left\{ P_{GG}(z)~
\left[ 1 +  { 1 \over 2 } f_{G} (xz, \tau) \right]
\left[ 1 +  { 1 \over 2 } f_{G} \left( x \left(
1 - z \right), \tau \right) \right]  \right. \nonumber \\
& +&  {1 \over 2}  \sum_{j=1}^{n_{f}} \sum_{\lambda=+,-}
P_{qG}(z)~ \left\{
\left[ 1 -  f_{j}^{\lambda} (xz, \tau)\right]
\left[ 1 -  \bar{f}_{j}^{-\lambda} \left( x \left(
1 -  z \right), \tau \right) \right]  \right. \nonumber \\
& + & \left. \left.  \left[ 1 -
\bar{f}_{j}^{\lambda} (xz, \tau)\right]
\left[ 1 -  f_{j}^{-\lambda} \left( x \left(
1 -  z \right), \tau \right) \right] \right\} \right\}~~~.
\label{eq:gluoni}
\enqa

Several comments can be made on the above expressions. First of all
we notice that the inverse decay processes, the ones
contained in the $C^{-}_{i}$ collisional integral,
contribute to the scaling behaviour of parton
distribution functions with terms quadratic in the $p_{A} (x,\tau)$
at least.
This is a consequence of our interpretation of A-P equations as
{\it transport equations}.
As already shown these terms vanish in the limit of a very dilute system.

The generalized equations predict also
a different, more complicated, evolution for momenta. By taking Mellin
transform of both sides of (\ref{eq:quark}) and
(\ref{eq:gluoni}), in fact, one sees that the standard scaling behaviour
should be corrected by terms quadratic and cubic in distribution functions,
which are not simply products of momenta of quarks and gluon densities.

Finally, as for the standard A-P equations, the scaling behaviour for
unpolarized quark distributions can be obtained by simply considering the
sum $q_{j}(x,\tau) = q_{j+}(x, \tau) + q_{j-}(x,\tau)$ (the same holds
for antiquarks).
Notice, however, that since the introduction of final state statistical
factors spoils the linearity of the equations, the evolution of $q_{j}(x,\tau)$
will depend on both the polarized distribution functions and not simply on
their sum.

We are now at some concluding remarks. We have stressed the point that, as
some experimental results suggest, the Fermi or Bose nature of partons could
sensibly manifest itself in observable quantities in deep inelastic scattering
on nucleons. This idea already successfully applied in \cite{buc} and
\cite{bour} mostly motivates our paper.
In particular it seems to us quite natural that quantum statistics
may modify the scaling behaviour of parton distribution functions
for rather small $x$ and high $Q^2$; in this region the sea
becomes dominant and thus bremsstr\"{a}hlung processes, responsible at
leading-log level for scaling breaking, are likely supposed to occur in
presence of a
{\it gas} of partons. In this case Pauli blocking and gluon stimulated
emission play, in general, a sensible role in parton distribution evolution.

We have introduced both this statistical effects
obtaining the generalized scaling, starting from the observation that
a quite close and intriguing analogy seems to hold between A-P equations
and a set of Boltzmann transport equations for partons.
In our approach to nucleons as {\it statistical} systems, this fact
is of course welcome and expected.
Pursuing this formal analogy it seems to us very fascinating the fact
that the scale variable $Q^2$ can be interpreted in some sense
as a {\it time} parameter.
The physical significance of this point, if any, should be deeper understood.
It is also rather interesting to stress that, from this point of view,
one would naturally expect, in the spirit of Boltzmann $H$ theorem,
that the normalized parton distributions $f_{j\lambda}(x,\tau)$,
$\bar{f}_{j\lambda}(x,\tau)$ and $f_{G}(x,\tau)$
should approach stationary Fermi and Bose  expressions as
$Q^2$ increases. Remarkably, these conclusions seem to agree with the
phenomenological results obtained in \cite{bour}.

\bigskip

We are much indebted with Prof. Franco Buccella, which we are
pleased to thank for
indefatigably encouraging this work and for his valuable comments.
We would like also to thank Prof. E. W. Kolb, Dr. V. A. Bedniakov and Dr. S. G.
Kovalenko for useful discussions. This work was supported in part by the DOE
and by the NASA (NAGW-2381) at Fermilab.

\end{document}